\documentclass[aps,prl,floatfix,twocolumn,showpacs]{revtex4}
\usepackage{amssymb}
\usepackage{amsmath}
\usepackage{graphicx}

\begin{document}

\title{Spectroscopy of non-local superconducting correlations in a double quantum dot.}
\author{L.G. Herrmann$^{1,5}$, P. Burset$^{2}$, W.J. Herrera$^{3}$, F. Portier$^{4}$, P. Roche$^{4}$, C. Strunk$^{5}$, A. Levy Yeyati$^{2}$ and T. Kontos$^{1}$\footnote{To whom correspondence should be addressed: kontos@lpa.ens.fr}}
\affiliation{$^{1}$Laboratoire Pierre Aigrain, Ecole Normale Sup\'erieure, CNRS UMR 8551, Laboratoire associ\'e aux universit\'es Pierre et Marie Curie et Denis Diderot, 24, rue Lhomond, 75231 Paris Cedex 05, France\\
$^{2}$Departamento de F\'{\i}sica Te\'{o}rica de la Materia Condensada C-V, Universidad Aut\'onoma de Madrid, E-28049 Madrid, Spain.\\
$^{3}$Departamento de F\'{\i}sica, Universidad Nacional de Colombia, Bogot\'{a}, Colombia.\\
$^{4}$Service de physique de l'\'etat Condens\'e, CEA, 91192
Gif-sur-Yvette, France.\\
$^{5}$ Institut f\"ur experimentelle und angewandte Physik ,
Universit\"at Regensburg, Universit\"atsstr.31,  93040 Regensburg,
Germany.}

\pacs{73.23.-b,73.63.Fg}

\begin{abstract}
We investigate non-linear transport in a double quantum dot connected to two normal electrodes and a central superconducting finger. By this means, we perform a transport spectroscopy of such a system which implements a Cooper pair splitter. The non-linear conductance exhibits strong subgap features which can be associated with the coherence of the injected Cooper pairs. Our findings are well accounted for by the recently developed microscopic theory of Cooper pairs splitters made in SWNTs.
\end{abstract}

\date{\today}
\maketitle

Combining the quantum coherence arising from superconductivity and the tunability of transport
in quantum dot circuits is interesting from both a fundamental and a practical point of view. On the one hand, it allows to investigate phenomena
occuring in conventional superconducting/normal metal structures with exquisite details thanks to the tunability of the physically
relevant parameters \cite{Pillet:10,Mason:10}. On the other hand, they offer an interesting platform for studying the interplay of electron-electron interaction and superconductivity in confined
systems (see \cite{LevyYeyati:11} for a recent review).

Recently, it has been shown experimentally that the basic elements of superconductors, i.e. the Cooper pairs, could be split in a controlled fashion in double quantum dot structures made in single wall
carbon nanotubes \cite{Herrmann:10} or in InAs nanowires\cite{Hofstetter:09}. If coherence of the emitted entangled states could be demonstrated, these devices would constitute important building blocks for implementing quantum optics-like experiments with electronic systems.

\begin{figure}[!hpth]
\centering\includegraphics[height=0.6\linewidth,angle=0]{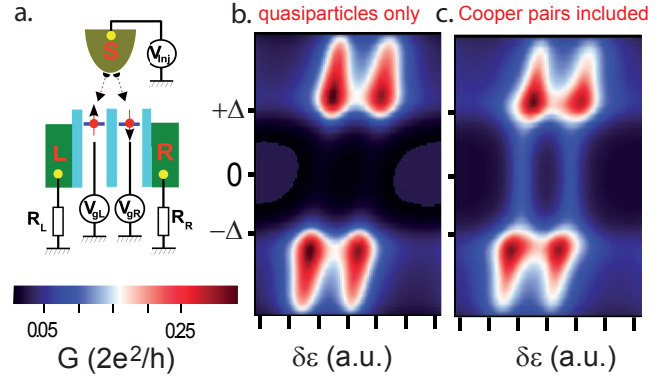}
\caption{a. Schematics of the two levels of the double quantum dot being close to resonance where the splitting action of the Cooper pair splitter is enhanced. b. Expected conductance map when the CPS is close to situation a. for strong tunneling to the normal contacts (in colorscale plot) when only quasiparticle tunneling is included. The subgap conductance is vanishingly small. c. Expected conductance map when the CPS is close to situation a. for strong tunneling to the normal contacts (in colorscale plot) when both Andreev tunneling and quasiparticle tunneling are included. The latter corresponds to the experimental situation depicted here and displays the characteristic subgap conductance ridges.}%
\label{Figure1:device}%
\end{figure}

So far though, a spectroscopic observation of split Cooper pairs emitted by a superconductor onto a double quantum dot system is lacking \cite{Hofstetter:09,Herrmann:10,Hofstetter:11,Schindele:12}. Before proceeding to more sophisticated experiments like noise correlations or photon emission\cite{Martin:96,Borlin:02,Samuelsson:02,Recher:01,Cottet:12}, it is important to establish on solid grounds that the split Cooper pairs maintain their coherence for at least their dwell time in the double dot. Here, we show that the non-linear regime is particularly relevant for implementing a spectroscopy of these non-local superconducting correlations. It allows to demonstrate the existence of subgap states which arise from the crossed Andreev reflections (equivalent to Cooper pair splitting) and the partial confinement in each of the two dots. This shows that the split Cooper pairs are coherent even when injected in the double dot, a result which could not be inferred from the linear transport regime. Finally, our findings are well accounted for by the recently developed microscopic theory of Cooper pairs splitters made in SWNTs\cite{Burset:11} and give an optimum bias for injecting split Cooper pairs over quasiparticles.

Recent spectroscopic studies of carbon nanostructures coupled to superconducting electrodes have allowed a direct test of the presence of Andreev bound states in these systems \cite{Pillet:10,Mason:10}. These states are naturally tunable thanks to quantum interference. Likewise, non-local Andreev states are expected to appear in a CPS geometry which is sensitive to the electron energy levels in each dot. These states should manifest in the transport properties of the device when the dot levels lie close to the Fermi energy of the normal electrodes, as originally proposed by Recher et al \cite{Recher:01}. Following the same spirit of our study in the linear regime\cite{Herrmann:10}, we consider the levels far from or close to the Fermi energy of the normal electrodes as the two extremal situations. For these two situations, one expects the Andreev states to be radically different.

Our setup, however, differs substantially from the recent spectroscopic studies \cite{Pillet:10,Mason:10} in at least two respects. First, we probe non-local Andreev states when the CPS is operated close to the situation of figure \ref{Figure1:device}a, yielding similar current to the left and the right arm of the CPS\cite{Herrmann:10}. Second, the CPS is connected to two normal electrodes which significantly broaden the Andreev levels. When the coupling to the normal leads is weak enough i.e. smaller than the energy gap $\Delta$ of the superconducting finger, the expected spectroscopy should be similar to that of Andreev levels exhibiting a minigap when the hybridized double dot levels cross the Fermi energy. However, when the coupling to the normal leads is bigger than $\Delta$ which corresponds to our experimental situation \cite{Herrmann:10}, the resulting transport spectroscopy displays broad features close to the two resonances. The presence of the Andreev states is revealed qualitatively by two parallel conductance ridges in the subgap region which correspond to broadened Andreev resonances, as shown in \ref{Figure1:device}c. The conductance map including Andreev processes  shown in \ref{Figure1:device}c is qualitatively distinct from that of pure quasiparticle tunneling shown in \ref{Figure1:device}b.

\begin{figure}[!hpth]
\centering\includegraphics[height=0.99\linewidth,angle=0]{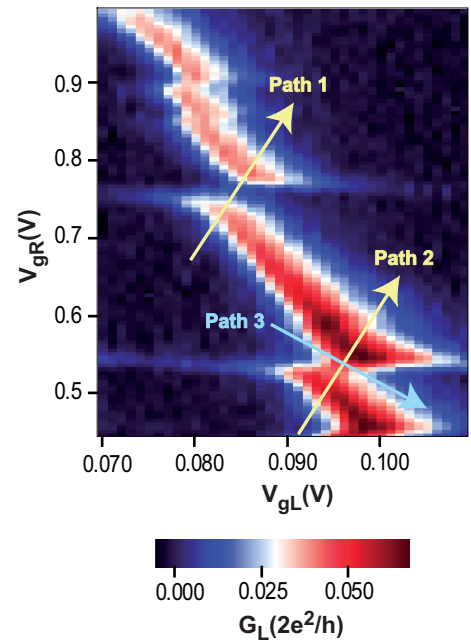}
\caption{Stability diagram of the CPS as measured by injecting current from the superconducting finger in the linear regime and measuring the conductance to the left(L) normal electrode. The three paths for which the non-linear spectroscopy is studied are indicated by yellow and blue arrows.}%
\label{Figure2:honeycomb}%
\end{figure}

Our measurements are performed on a device based on a single wall carbon nanotube connected to an Al/Pd superconducting electrodes and two Au/Ti normal electrodes (details on the fabrication can be found in \cite{Herrmann:10}). The temperature is $120 mK$ throughout the paper. We present only measurements of the conductance of the left arm $G_L$ of the CPS since the conductance of the right arm corresponds essentially to the $G_L$ up to a scale factor. The linear stability diagram for the gate region studied here is displayed in figure \ref{Figure2:honeycomb}. The characteristic avoided crossings of levels is observed. In order to investigate the non-linear spectroscopy of the CPS, it is convenient to follow the paths sketched with the blue and yellow arrows. Paths 1 and 2 correspond to a situation where the left level $\epsilon_L$ and the right level $\epsilon_R$ are moved in the same direction ($\delta \epsilon_L \approx \delta \epsilon_R$). Path 3 corresponds to a situation where the left and right levels are moved in opposite directions ($\delta \epsilon_L \approx -\delta \epsilon_R$).

\begin{figure}[!hpth]
\centering\includegraphics[height=1.0\linewidth,angle=0]{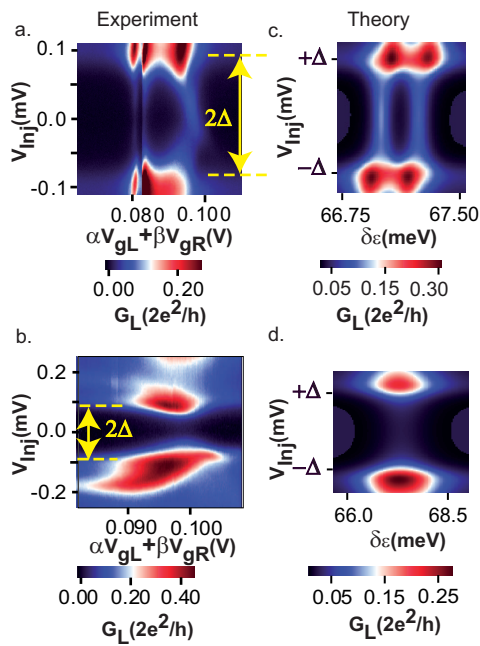}
\caption{a. Zoom on the non-linear spectroscopy of the CPS in color scale of $G_L$ along path 2. Characteristic subgap conductance ridges resulting from the non-local Andreev levels are observed. b. Zoom on the non-linear spectroscopy of the CPS in color scale of $G_L$ along path 3. c. Theoretical color scale plot obtained for path 2 using the microscopic theory of the CPS. d. Theoretical color scale plot obtained for path 3 using the microscopic theory of the CPS. }%
\label{Figure4:egg}%
\end{figure}

Figure \ref{Figure4:egg}a and b display a zoom on the spectroscopy of the CPS when the side gate voltages are swept along paths 2 and 3 (respectively). Note that this corresponds to the second anticrossing indicated in figure \ref{Figure2:honeycomb}. Path 2 corresponds to the path used for our previous Cooper pair splitting analysis in the linear regime \cite{Herrmann:10}. The energy gap in the conductance is clearly visible on the left side and the right side of each color scale plot. A clear dispersion of the inner edges of the conductance peaks is observed. Strikingly, we observe the characteristic spectroscopy of the Andreev states shown in figure \ref{Figure1:device}c. The non-linear spectroscopy along path 3 is also informative since it corresponds to level shifts which preserve the Andreev reflection since the energy is reverted between one dot and the other. In that case shown in figure \ref{Figure4:egg}b, the subgap conductance disappears and a pinching of the gap edges is observed.

For analyzing these experimental results, we follow  Ref. \cite{Burset:11} and use a tight-binding (TB) model in the usual nearest-neighbor approximation with a hoping parameter $t_g\simeq 2.7eV$ for describing zigzag single-walled carbon nanotubes (SWNT). We consider metallic nanotubes with radius $R\sim 0.47$nm, which have a narrow curvature gap $E_{curv}\sim 45$meV. A SWNT of length $600$nm is contacted by normal metallic leads at its ends and by a central superconducting lead of width $90$nm.
The lateral metallic leads are modeled by ideal one-dimensional channels coupled to the ends of the tube with tunneling rates $\Gamma_{L}\simeq 0.1t_g$ and $\Gamma_{R}\simeq 0.03t_g$.
We consider the central part to be contacted by an Al electrode. According to the {\it ab-initio} calculations of Ref. \cite{Kuemmeth:08} this electrode produces in the normal state a {\it n}-doping effect, leading to a shift of the tube bands $E_{FS}\sim -0.5$eV, for an ideal interface.
This structure exhibits an anticrossing pattern in the gate plane similar to the experimental one when the potential profile in the normal regions ($V_{gL}$ and $V_{gR}$) is tuned to populate the valence bands ($V_{gL},V_{gR}>E_{curv}$, i.e. in the {\it p-n-p'} regime).
In the superconducting state, it induces a gap $\Delta_i \sim 0.1$meV and we estimate the tunneling rate between the SWNT and this electrode to be of the same order. $\Gamma_s$ denotes the effective coupling of the resonant levels to the SC finger with an energy gap $\Delta$. The coupling to the central electrode is assumed to be smooth enough on the atomic scale to neglect inter-valley scattering.
However, for small radius the spin-valley degeneracy is broken by spin-orbit interactions. In the results of figure \ref{Figure4:egg}c and d, only the crossing of one of these spin-valley resonances for each dot is shown. Notice that in the regime we are considering, where Kondo correlations can be neglected and $U >>\Gamma_{L,R} > \Delta$, the main effect of Coulomb interactions is to open a gap between consecutive spin-valley resonances. The behavior found in the microscopic calculations is reproduced qualitatively with a minimal model in which a single level is used to describe each dot as in the one introduced in Ref. \cite{Herrmann:10}. The parameter $\Gamma_{12}$ which describes the coupling between the two dots is the one controlling the separation of Andreev resonances along path 2. The rest of the parameters follow the same hierarchy of energy scales used in Ref. \cite{Herrmann:10}.

Figure \ref{Figure4:egg}c and d display the color scale plots of the calculated conductance $G_L$ for the realistic parameters given above. While in Ref. \cite{Burset:11} only the contribution of subgap processes (i.e. local and non-local Andreev reflections) to the conductance were taken into account, in the present results we add the contribution of single quasiparticle processes which give the dominant contribution for $eV > \Delta$. Due to the finite temperature these processes also contribute for $eV < \Delta$ but their effect is not significant for $T=120 mK$. We find a good agreement with the experiment for the two paths, both for the shape of the dispersion of the Andreev levels and for the amplitude of variations of the conductance. The observed dispersion of the Andreev levels is not fully reproduced within our minimal model. This may originate from, e.g., an energy dependence of the effective tunneling rates from the double dot to the leads. The theoretical analysis reveals that the gate dependence of the subgap features along path 2 is a clear signature of the dominance of the non local Andreev processes compared to single quasiparticle processes. When temperature is increased and single quasiparticle processes are enhanced such structure becomes gradually tilted and less pronounced.

\begin{figure}[!hpth]
\centering\includegraphics[height=0.8\linewidth,angle=0]{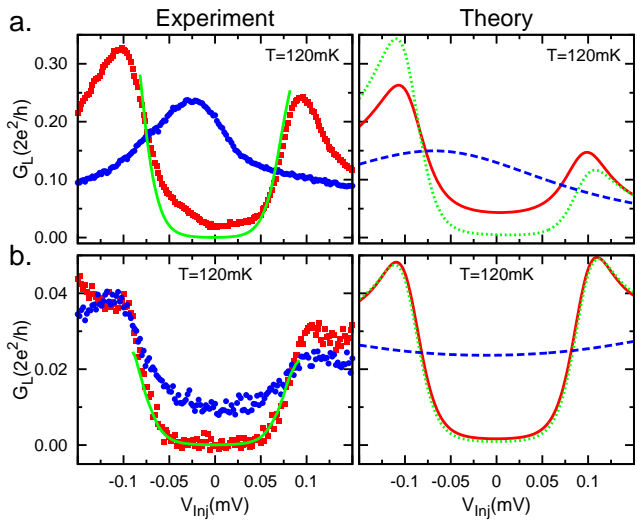}
\caption{ a. Left panel : Cut of the measured non-linear conductance of the CPS along path 1, at resonance, taken at $0mT$ (in red squares) and at $90mT$ (in blue circles). In solid green line, the convolution of BCS density of state with a gap of $85 \mu eV$ and the \textit{measured} normal state differential conductance at $120 mK$. Right panel : Corresponding theoretical plots calculated with the microscopic theory. The superconducting state curve is in solid red lines and the normal state curves are in blue dashed lines. We show the calculation to lowest order in $\Gamma_s$ in green dashed lines. b. Same plots as for panel a. off resonance.}%
\label{Figure4b:eggcut}%
\end{figure}

The experimental findings presented so far show qualitatively the existence of non-local superconducting correlations on the double quantum dot. We now provide quantitative evidence for the existence of the non-local superconducting correlations.
Figure \ref{Figure4b:eggcut}a and b left panels displays a cut of the non-linear spectroscopy of the CPS taken along path 1 at resonance and off resonance respectively. For both cases, the measurement are taken at $0mT$ and $89mT$, as shown in red squares and blue circles respectively. These two values of external magnetic field are used to define the superconducting and the normal state of the CPS respectively.

Outside the anticrossing points, a gap $\Delta \approx 85 \mu V$ opens around zero bias, consistently with figure \ref{Figure4:egg}a and b. BCS-like quasiparticle peaks surround the gap. For energies below that of the quasiparticle peaks, the measured differential conductance in the superconducting state is quantitatively fitted to a convolution of the BCS density of state with a gap of $85 \mu eV$ and the \textit{measured} normal state differential conductance at $120 mK$ \cite{BCScurve}, as shown by the green solid line in figure \ref{Figure4b:eggcut}b left panel. Importantly, no gap is found at the resonances of an anti-crossing as shown in figure \ref{Figure4b:eggcut}a left panel, in red squares. In that case, the normal state differential conductance displays a peak (in blue circles). Here, the quasiparticle peaks get broadened and the change of shape of the low energy part of the differential conductance indicates the presence of midgap states. Strikingly, the convolution of the BCS density of states with the normal state conductance, in green solid line does not account for the level of the measured subgap conductance. This shows that the measured subgap conductance in the superconducting state cannot be simply accounted for by electrical transport of the series of a superconductor and a double quantum dot. The excess conductance measured at the anti-crossings can only be explained by the existence of superconducting correlations on the double quantum dot, which are non-local here due to the nature of the device. This behaviour is very well reproduced by the microscopic theory, shown in the right panel for each case. The superconducting state curve is in solid red lines and the normal state curves are in blue dashed lines \cite{Kondonote}. Here, for comparison, we show the calculation to lowest order in $\Gamma_s$ in green dashed lines which is equivalent to the convolution of the BCS density of states with the normal state conductance. Note that the observed behaviour allows us to determine an optimum for the maximizing of the DC current of split Cooper pairs over the quasiparticle current at finite bias. We find $V_{inj} \approx \Delta/2$ which corresponds to the maximum of the Cooper pair current measured by the differential conductance over the expected quasiparticle conductance.

In conclusion, we have investigated the operation of a Cooper pair splitter in the non-linear regime. The conductance of each arm of the beam splitter implements a spectroscopy of the system. We have identified non-local superconducting correlations inside the double quantum dot connected to the normal leads and the central superconducting electrode. This shows that Cooper pairs can survive once injected in the double quantum dot and gives an optimum of $V_{inj} \approx \Delta/2$ for maximizing the split Cooper pair current over the quasiparticle current.

\begin{acknowledgments}
We thank A. Cottet for a critical reading of the manuscript and
illuminating discussions. The devices have been made within the consortium Salle Blanche Paris Centre. This work is supported by the SFB 689 of the Deutsche Forschungsgemeinschaft, by the ANR contracts DOCFLUC, HYFONT and SPINLOC, the Spanish MICINN under
contract FIS2008-04209 and the DFH-UFA and DAAD mobility grants and the EU-FP7 project SE2ND[271554].
\end{acknowledgments}


\begin{thebibliography}{99}
\bibitem{BCScurve} The convolution procedure is reliable only at low energies due to the lack of high bias data for the normal state conductance.
\bibitem{Kondonote}Note that the observed peak in the normal state conductance is narrower than the one obtained in the calculations. This might indicate the emergence of a Kondo resonance.
\end{thebibliography}
\end{document}